\documentclass[pra,showpacs,twocolumn]{revtex4}
\usepackage{amsmath,amsfonts,graphicx}
\usepackage[amssymb]{SIunits}
\usepackage{color}

\newcommand{\ket}[1]{|#1\rangle}
\newcommand{\bra}[1]{\langle#1|}
\newcommand{\bracket}[2]{\langle#1|#2\rangle}
\renewcommand{\vec}{\textbf}

\begin{document}

\title{Strange behavior of the relativistic Einstein-Podolsky-Rosen
  correlations} 

\author{Pawe{\l}{} Caban}\email{P.Caban@merlin.phys.uni.lodz.pl}
\author{Jakub Rembieli{\'n}ski}\email{jaremb@uni.lodz.pl}
\author{Marta W{\l}odarczyk}\email{marta.wlodarczyk@gmail.com}

\affiliation{Department of Theoretical Physics, University of Lodz\\
Pomorska 149/153, 90-236 {\L}{\'o}d{\'z}, Poland}

\date{\today}

\begin{abstract}
We show that configurations exist in which the correlation
functions and the degree of violation of Bell-type inequalities
in the relativistic Einstein-Podolsky-Rosen (EPR) experiment
have local extrema for some values of the
velocities of the EPR particles. Moreover, this strange behavior can
be observed 
for both discussed relativistic spin operators and for spin-1/2 as
well as spin-1 particles. 
\end{abstract}


\maketitle

Until recently, almost all  papers concerning quantum-information
processing were based on the non-relativistic quantum 
mechanics. On the other hand the present technological
possibilities speed up the investigation of the relativistic aspects
of the quantum Einstein-Podolsky-Rosen (EPR) correlations.

The aim of this paper is to report some strange behavior of the
relativistic EPR  
correlation functions. We show that the correlation function,
which in the relativistic case depends on the particle momenta, for
some fixed configurations has local extrema.
Such a bechvior has not been reported in the previous works
\cite{ALMH2003,TU2003_1}.
Such extrema can be
observed for both spin-1/2 and spin-1 particles and for two different
choices of the relativistic spin operator. 
This suggests that the discussed effect is a general property of the
relativistic correlation functions.
We also show that
relativistic quantum correlations are stronger than nonrelativistic
ones for a variety of configurations. Consequently, in such
configurations Bell inequalities are more strongly violated by
relativistic correlations than by nonrelativistic ones.

An appropriate treatment of the EPR experiment
is hindered by very serious
theoretical and interpretational difficulties concerning the
relativistic quantum mechanics.
One of the
most frustrating problems is the lack of the
Lorentz-covariant notion of localizability in the relativistic quantum
mechanics. 
The position operator is needed not only to take
into consideration the finite size of the detectors but also
it is directly related with the
definition and form of spin operator.

The most familiar choice of the position operator for a massive
particle is the Newton-Wigner operator \cite{cab_NW1949}
 \begin{equation}
 \hat{\vec{Q}}_{NW}   = - \frac{1}{2} \Big[
 \frac{1}{\hat{P}^0}\hat{\vec{K}} + \hat{\vec{K}}
 \frac{1}{\hat{P}^0} \Big] - 
 \frac{\hat{\vec{P}} \times \hat{\vec{W}}}{m \hat{P}^0 (m + \hat{P}^0 )}, 
 \end{equation}
where  $\hat{P}^0$, $\hat{\vec{P}}$ are the four-momentum operators,
$m$ the particle mass, $\hat{\vec{K}}$  is the Lorentz boosts
generator, and  $\hat{\vec{W}}= \hat{P}^0 \hat{\vec{J}} + \hat{\vec{P}}
\times \hat{\vec{K}}$  
is the space part of the Pauli-Lubanski four-vector 
$\hat{W}^{\mu} = \frac{1}{2}\epsilon^{\nu\gamma\delta\mu}
    \hat{P}_{\nu}\hat{J}_{\gamma\delta}$, $\hat{\vec{J}}$ is the 
total angular momentum operator, and $\hat{K}^i=\hat{J}^{0i}$,
$\hat{J}^i=\varepsilon^{ijk}\hat{J}^{jk}$. The Newton-Wigner operator
forms a 
vector with commuting, self-adjoint  components and is defined for
arbitrary spin. Another popular choice of the  position operator  is
\cite{cab_Bacry1988}
 \begin{equation}
 \hat{\vec{Q}}_{CM} = - \frac{1}{2} 
  \Big[ \frac{1}{\hat{P}^0}\hat{\vec{K}} + \hat{\vec{K}}
  \frac{1}{\hat{P}^0} \Big]
 \end{equation}
interpreted also as the center-of-mass position operator. For spinning
particles components of this operator do not commute.
Unfortunately, both operators do not form any autonomous geometrical
object with a covariant transformation law. 

Now, for observers
in the same inertial frame spin is defined as a
difference between total angular momentum (which is well defined as
the generator of the rotations) and the orbital angular momentum
$\hat{\vec{L}}= \hat{\vec{Q}} \times \hat{\vec{P}}$: 
 \begin{equation}
 \hat{\vec{S}}= \hat{\vec{J}} - \hat{\vec{Q}} \times \hat{\vec{P}}.
 \end{equation}
However,  to define
the orbital angular momentum  $\hat{\vec{L}}$ we should know the
relativistic position operator 
$\hat{\vec{Q}}$. The lack of a generally accepted position
operator results in ambiguities in the definition of relativistic spin
operator. 
In particular, Newton-Wigner operator $\hat{\vec{Q}}_{NW}$
leads to the following spin observable:
 \begin{equation}
    \label{eq:spin_operator_our_def}
    \hat{\vec{S}}_{NW} = \frac{1}{m} \left(\hat{\vec{W}}
    -\hat{W}^0\frac{{\hat{\vec{P}}}}{\hat{P}^0+m}\right),
 \end{equation}
which satisfies usual spin algebra [su(2) Lie algebra].
$\hat{\vec{S}}_{NW}$ is the only
axial-vector operator being  linear function of  the
Pauli-Lubanski four-vector \cite{cab_BLT1969}. 
However, such spin operator is neither an autonomous 
geometrical object under Lorentz transformations nor even a part of an
irreducible object. 

For the position operator $\hat{\vec{Q}}_{CM}$ the corresponding  spin
observable takes the form
 \begin{equation}
  \hat{\vec{S}}_{CM}= \frac{\hat{\vec{W}}}{\hat{P}^0}.
 \label{eq:spin_CM}
 \end{equation}
Unfortunately, components of this operator do not form the spin
algebra. Moreover their eigenvalues $\lambda_i$ are momentum-dependent  i.e. 
$\lambda_i = \lambda \sqrt{[ m^2 +(k^i)^2]}/k^0$ , $\lambda = -s,
-s+1,\dots,s$. Furthermore, in contrast to the operator
$\hat{\vec{S}}_{NW}^2$, the operator $\hat{\vec{S}}_{CM}^{2}$ does not
reduce to the relativistic spin-square operator $-W^\mu W_\mu/m^2$
equal to $s(s+1)\openone$ in an unitary irreducible representation of
the Poincar\'e group.
Therefore the operator (\ref{eq:spin_CM})
cannot be treated as a proper spin observable. 
For this reason Czachor
\cite{Czachor1997_1}, and following him a number of authors (see,
e.g., Refs.~\cite{ALMH2003,LY2004,Moradi2008}),  
used the normalized operator (\ref{eq:spin_CM}). 
One can easily show that the spin observable used in
Ref.~\cite{Czachor1997_1} can be cast in the following form:
 \begin{equation}
  \hat{\vec{S}}(\vec{a})= 
  \frac{\vec{a}\cdot
  \hat{\vec{W}}}{\sqrt{m^2+(\vec{a}\cdot\hat{\vec{P}})^2}}.
 \label{eq:Czachor_projection_def}
 \end{equation}
This operator has a proper spectrum;
however, it cannot be treated as a 
projection of any spin observable on the direction $\vec{a}$,
$|\vec{a}|=1$, because it is a nonlinear function of $\vec{a}$.  
In the following we will compare
EPR correlations and Bell-type inequalities
obtained with help  of the spin
operator (\ref{eq:spin_operator_our_def}) and the
operator (\ref{eq:Czachor_projection_def}).

Let us consider two distant observers Alice and Bob in the same
inertial  frame, sharing a pair of particles with sharp momenta in a
two-particle state  $\ket{\Psi}$.
We take into account only such
measurements in which Alice and Bob register one particle each.
Without loss of generality we can assume that particles are
distinguishable and Alice registers the particle with momentum equal
to $\vec{k}$ and Bob the particle with momentum equal to $\vec{p}$.
Now let Alice measure spin component of her particle in direction
$\vec{a}$ and Bob spin component of his particle in direction
$\vec{b}$, where $|\vec{a}|=|\vec{b}|=1$. Their observables are
$(\vec{a}\cdot\hat{\vec{S}}_{NW})\otimes\openone$
and $\openone\otimes(\vec{b}\cdot\hat{\vec{S}}_{NW})$, when one uses
the spin operator $\hat{\vec{S}}_{NW}$ defined in
Eq.~(\ref{eq:spin_operator_our_def}), or
$\hat{\vec{S}}(\vec{a})\otimes\openone$ and
$\openone\otimes\hat{\vec{S}}(\vec{b})$ when one uses the operator
(\ref{eq:Czachor_projection_def}). 
Consequently, the normalized
correlation function in the EPR-type experiment has the form
 \begin{equation}
 \label{eq:correl_func_NW_def}
 \mathcal{C}^{\Psi}(\vec{a},\vec{b}) = 
 \frac{\bra{\Psi} (\vec{a}\cdot\hat{\vec{S}}_{NW})
 \otimes (\vec{b}\cdot\hat{\vec{S}}_{NW})
 \ket{\Psi}}{s^2\bracket{\Psi}{\Psi}},
 \end{equation}
for the spin operator $\hat{\vec{S}}_{NW}$ or
 \begin{equation}
 \label{eq:correl_func_CM_def}
 \mathcal{C}^{\Psi}_{\textsf{Cz}}(\vec{a},\vec{b}) = 
 \frac{\bra{\Psi} \hat{\vec{S}}(\vec{a}) \otimes \hat{\vec{S}}(\vec{b})
 \ket{\Psi}}{s^2\bracket{\Psi}{\Psi}},
 \end{equation}
for the operator (\ref{eq:Czachor_projection_def}) proposed by Czachor.

In this paper
we will discuss EPR correlations in two-particle states which
are singlets of the Lorentz group. 
Some of the formulas presented in this paper have been obtained in
our previous works but we include them to the present paper to make it
self-consistent.
Let us remind first the notation concerning one-particle states.
For the particle with mass $m$ and spin $s$ space of states
is spanned by the
four-momentum eigenvectors $\ket{k,m,s,\sigma}$.
These vectors are normalized covariantly.
The action of the Lorentz transformation $\Lambda$ on the vector
$\ket{k,m,s,\sigma}$ is of the form
 \begin{equation}
 U(\Lambda)\ket{k,m,s,\sigma}=
 {\mathcal{D}}_{\lambda\sigma}^{s}(R(\Lambda,k)) \ket{\Lambda
 k,m,s,\lambda},  
 \label{eq:transf_states_standard}
 \end{equation}
where ${\mathcal{D}}^s$ is the matrix spin $s$ representation of the SO(3)
group, $R(\Lambda,k)=L_{\Lambda k}^{-1}\Lambda L_k$ is the Wigner
rotation, and $L_k$ designates the standard Lorentz boost defined by
the relations $L_k\tilde{k}=k$, $L_{\tilde{k}}=I$,
$\tilde{k}=(m,\vec{0})$. Throughout all the paper we will assume that
both EPR particles have mass $m$. Moreover, for fixed values of the
spin we will use the notation $\ket{k,\sigma}\equiv\ket{k,m,s,\sigma}$.

For $s=1/2$ pseudoscalar state of two particles with sharp momenta
was discussed in Refs.~\cite{CR2005,CR2006}. It has the following
form: 
 \begin{multline}
 \label{eq:state_fermions}
 \ket{\varphi(k,p)} =
 \frac{-i}{\sqrt{2}
 \sqrt{\left(1+\frac{k^0}{m}\right)\left(1+\frac{p^0}{m}\right)}}\\
 \times
 \Bigg\{  \Big[ 
 \openone\big(1+\frac{k^0+p^0}{m}+\frac{kp}{m^2}\big)
 -\frac{i(\vec{k}\times\vec{p})\cdot\boldsymbol{\sigma}}{m^2}
 \Big] \sigma_2
 \Bigg\}_{\sigma\lambda}\\
 \ket{k,\sigma}\otimes\ket{p,\lambda},
 \end{multline}
where $\sigma,\lambda=\pm\frac{1}{2}$,
$\boldsymbol{\sigma}=(\sigma_1,\sigma_2,\sigma_3)$, and $\sigma_i$ are
standard Pauli matrices. In the center-of-mass (c.m.) frame
[$p=k^\pi\equiv(k^0,-\vec{k})$] the state 
(\ref{eq:state_fermions}) is an ordinary singlet state.

Correlation function (\ref{eq:correl_func_NW_def}) for
spin-1/2 particles in the 
state (\ref{eq:state_fermions}) was calculated in
Ref.~\cite{CR2005} and it reads
 \begin{multline}
 \label{eq:correl_fermions_our_general}
 \mathcal{C}^{\varphi(k,p)}(\vec{a},\vec{b}) = -\vec{a}\cdot\vec{b} 
 + \frac{(\vec{k}\times\vec{p})}{m^2+kp}\cdot
 \bigg((\vec{a}\times\vec{b}) \\
   +\frac{(\vec{a}\cdot\vec{k})(\vec{b}\times\vec{p}) -
     (\vec{b}\cdot\vec{p})(\vec{a}\times\vec{k})}{(k^0+m)(p^0+m)}\bigg).
 \end{multline}
Notice that
in the c.m.\ frame the above correlation function is the same as in
the nonrelativistic case
$\mathcal{C}^{\varphi(k,k^\pi)}(\vec{a},\vec{b}) =
-\vec{a}\cdot\vec{b}$.

On the other hand, one can check that
correlation function (\ref{eq:correl_func_CM_def}) in the
state (\ref{eq:state_fermions}) has the form
 \begin{multline}
 \label{eq:correl_func_fermions_Czachor}
 \mathcal{C}^{\varphi(k,p)}_{\textsf{Cz}}(\vec{a},\vec{b}) 
 = \frac{m^2}{\sqrt{m^2+(\vec{a}\cdot\vec{k})^2}\sqrt{m^2+
     (\vec{b}\cdot\vec{p})^2}} \\
 \times \Big\{ -\vec{a}\cdot\vec{b} + 
 \frac{(\vec{a}\cdot\vec{k})(\vec{b}\cdot\vec{p})}{m^2}
 - \frac{[\vec{a}\cdot(\vec{k}+\vec{p})]
 [\vec{b}\cdot(\vec{k}+\vec{p})]}{m^2+kp} \Big\}.
 \end{multline}
In the c.m.\ frame it takes the form obtained by Czachor
\cite{Czachor1997_1}.

\begin{figure}
\centering
\includegraphics[width=1\columnwidth]{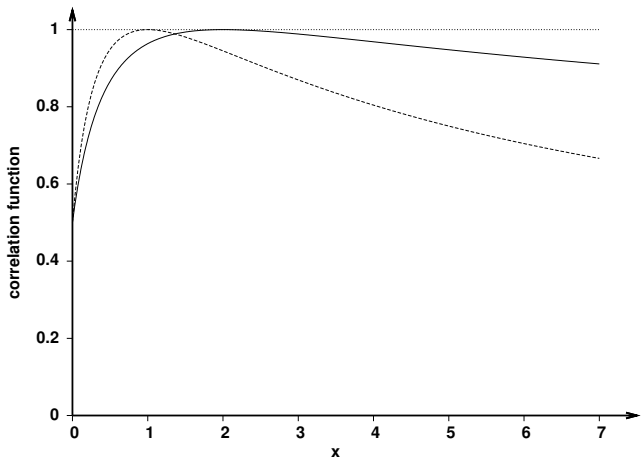}
\caption{The plot shows dependence of correlation functions
$\mathcal{C}^{\varphi(k,p)}(\vec{a},\vec{b})$
(solid line) and 
$\mathcal{C}^{\varphi(k,p)}_{\textsf{Cz}}(\vec{a},\vec{b})$
(dashed line) on $x$ for $k$ and $p$ given in
Eqs.~(\ref{seq:k_p_fermions}) and
$\vec{a}=(0,0,1)$,
$\vec{b}=(\frac{\sqrt{3}}{2},0,-\frac{1}{2})$.} 
\label{fig:fermions_correl}
\end{figure}

The unexpected behavior of the correlation functions
(\ref{eq:correl_fermions_our_general}) and
(\ref{eq:correl_func_fermions_Czachor}) can be observed in the cases
when observers are not in the c.m.\ frame of the pair of EPR
particles. 
As an example, let us consider the situation
when in Alice's and Bob's inertial frame
\begin{subequations}
 \begin{align}
 & k^\mu=m(\sqrt{4x+1},\sqrt{x},0,-\sqrt{3x}),\\
 & p^\mu=m(\sqrt{4x+1},-\sqrt{x},0,-\sqrt{3x}),
 \end{align}%
\label{seq:k_p_fermions}%
\end{subequations}%
where
 \begin{equation}
 x= \frac{{\textsf{W}}^2}{4m^2}-1.
 \label{eq:x_def}
 \end{equation}
Here $\textsf{W}$ denotes invariant total energy of the two-particle
system in the c.m.\ frame. 
In this case we can find such configurations in which both correlation
functions, (\ref{eq:correl_fermions_our_general}) and
(\ref{eq:correl_func_fermions_Czachor}), posses local extrema---see
Fig.~\ref{fig:fermions_correl}. 

\begin{figure}
\centering
\includegraphics[width=1\columnwidth]{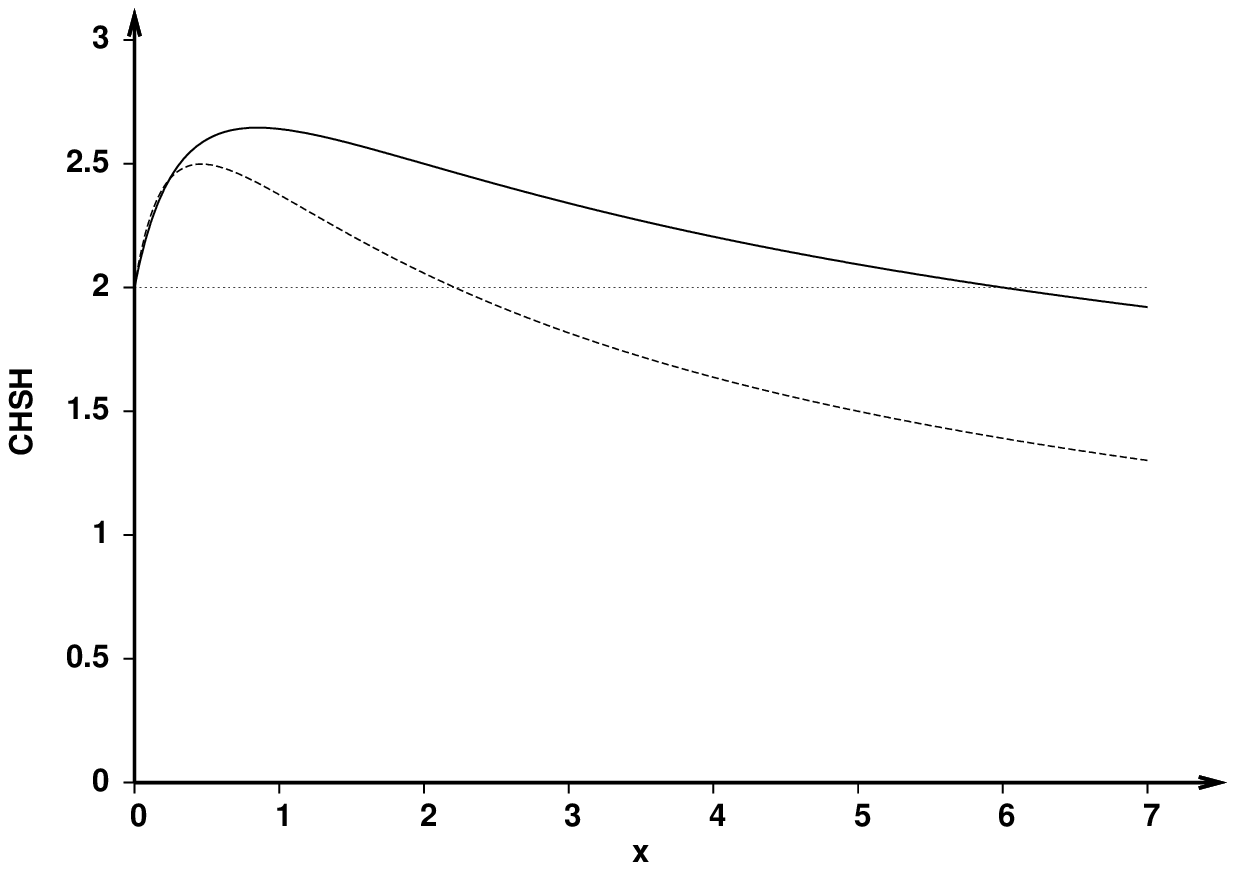}
\caption{The plot shows dependence of the left hand side of the CHSH
  inequality for $\mathcal{C}^{\varphi(k,p)}(\vec{a},\vec{b})$
(solid line) and for 
$\mathcal{C}^{\varphi(k,p)}_{\textsf{Cz}}(\vec{a},\vec{b})$
(dashed line)
on $x$ for $k$ and $p$ given in Eqs.~(\ref{seq:k_p_fermions}),
$\vec{a}=(0,0,1)$,
$\vec{b}=(0,0,1)$, 
$\vec{c}=(\frac{\sqrt{3}}{2},0,\frac{1}{2})$,
$\vec{d}=(\frac{\sqrt{3}}{2},0,\frac{1}{2})$.} 
\label{fig:fermions_CHSH_1}
\end{figure}

\begin{figure}
\centering
\includegraphics[width=1\columnwidth]{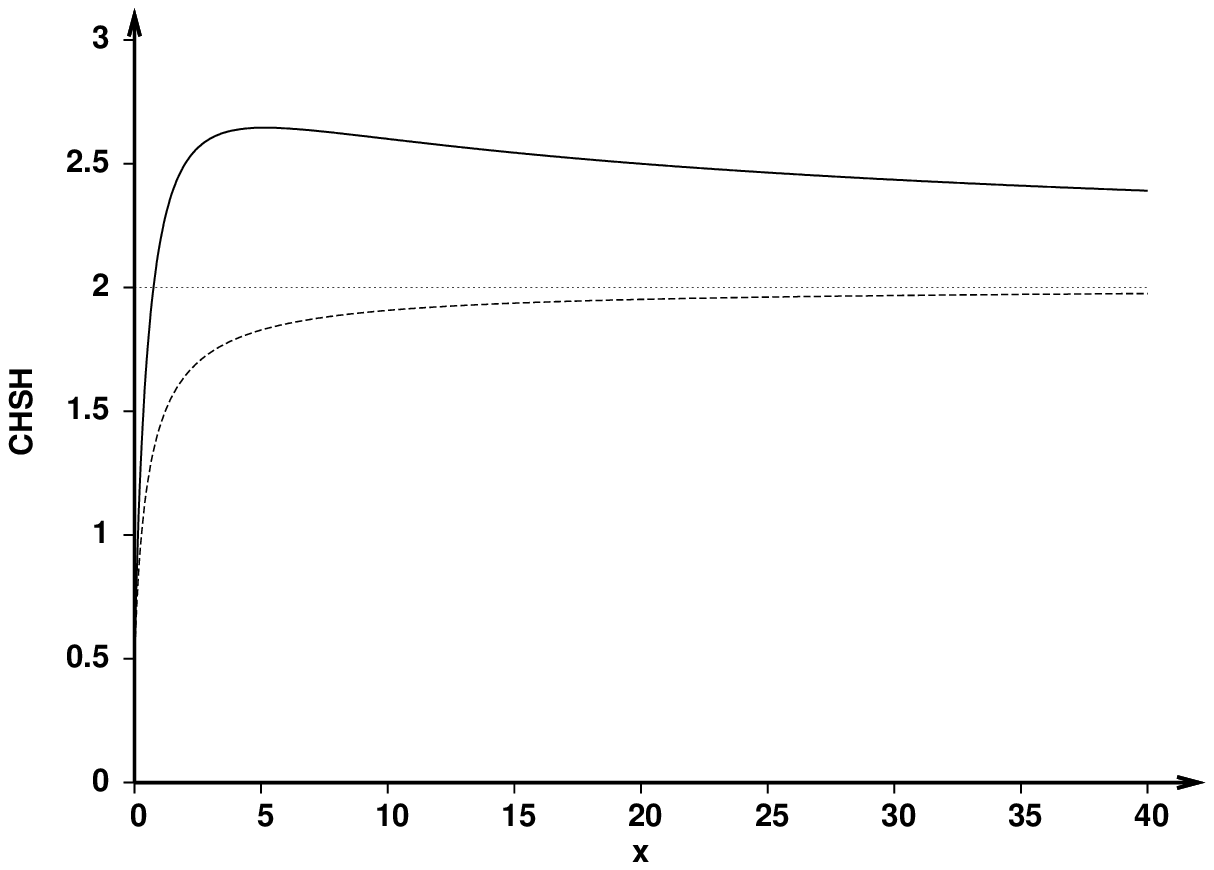}
\caption{The plot shows dependence of the left hand side of the CHSH
  inequality for $\mathcal{C}^{\varphi(k,p)}(\vec{a},\vec{b})$
(solid line) and for 
$\mathcal{C}^{\varphi(k,p)}_{\textsf{Cz}}(\vec{a},\vec{b})$
(dashed line)
on $x$ for $k$ and $p$ given in Eqs.~(\ref{seq:k_p_fermions}),
$\vec{a}=(0,0,1)$,
$\vec{b}=(0,0,1)$, 
$\vec{c}=(\frac{\sqrt{3}}{2},0,-\frac{1}{2})$,
$\vec{d}=(\frac{\sqrt{3}}{2},0,\frac{1}{2})$. } 
\label{fig:fermions_CHSH_3}
\end{figure}

This behavior of the correlation function has interesting physical
consequences. The violation of the Clauser-Horne-Shimony-Holt (CHSH)
inequality \cite{cab_Ballentine1998} 
 \begin{equation}
 \textsf{CHSH}=|\mathcal{C}(\vec{a},\vec{b}) - \mathcal{C}(\vec{a},\vec{d})
 + \mathcal{C}(\vec{c},\vec{b}) + \mathcal{C}(\vec{c},\vec{d})| \le2
 \label{eq:CHSH}
 \end{equation}
depends on the particle momenta and on the chosen spin operator. There
are configurations in which the quantity \textsf{CHSH} possesses local
maximum and exceeds 2 for 
both correlation functions,
$\mathcal{C}^{\varphi(k,p)}(\vec{a},\vec{b})$ and 
$\mathcal{C}^{\varphi(k,p)}_{\textsf{Cz}}(\vec{a},\vec{b})$---see
Fig.~\ref{fig:fermions_CHSH_1}. 
There are also such configurations in which the function 
$\mathcal{C}^{\varphi(k,p)}(\vec{a},\vec{b})$ violates
the CHSH inequality while the function 
$\mathcal{C}^{\varphi(k,p)}_{\textsf{Cz}}(\vec{a},\vec{b})$ does not
(Fig.~\ref{fig:fermions_CHSH_3}) and vice
versa. Notice that $x=0$ corresponds to the
nonrelativistic case. Therefore, in all configurations depicted in
Figs.~\ref{fig:fermions_CHSH_1}--\ref{fig:fermions_CHSH_3} the CHSH
inequality is not violated in the nonrelativistic case.

The situation is more surprising
for spin-$1$ particles. In this case we can observe similar phenomena
even in the c.m.\ frame. To see this let us consider the scalar
state of two spin-$1$ particles \cite{CRW2008}
 \begin{equation}
 \label{eq:state_bosons}
 \ket{\psi(k,p)}= \sum_{\sigma,\lambda=0,\pm1} 
 e^{\mu}_{\phantom{\mu}\sigma}(k) 
 e_{\mu\lambda}(p)
 \ket{k,\sigma}\otimes\ket{p,\lambda}, 
 \end{equation}
where the explicit form of the
amplitudes $e^{\mu}_{\phantom{\mu}\sigma}(k)$ can be found in
Ref.~\cite{CRW2008}. 

The correlation function $\mathcal{C}^{\Psi}(\vec{a},\vec{b})$
[Eq.~(\ref{eq:correl_func_NW_def})] in the
state (\ref{eq:state_bosons}) was calculated in
Ref.~\cite{CRW2008}.
In the c.m.\ frame this correlation function takes the form
 \begin{multline}
 \label{eq:correl_func_bosons_our_CMF}
 \mathcal{C}^{\psi(k,k^\pi)}(\vec{a},\vec{b}) = \\
    \frac{2}{2+(1+2x)^2}
    \big[-(1+2x)(\vec{a}\cdot\vec{b}) 
    +2x(\vec{a}\cdot\vec{n})(\vec{b}\cdot\vec{n})\big],
 \end{multline}
where $\vec{n}=\tfrac{\vec{k}}{|\vec{k}|}$ and $x$ is defined in
Eq.~(\ref{eq:x_def}). In the c.m.\ frame $x$ is connected to the
velocity of the particle via the relation $(v/c)^2=x/(x+1)$. Therefore
the correlation function (\ref{eq:correl_func_bosons_our_CMF}) depends
only on the velocity of the particles, not on its mass.

\begin{figure}
\centering
\includegraphics[width=1\columnwidth]{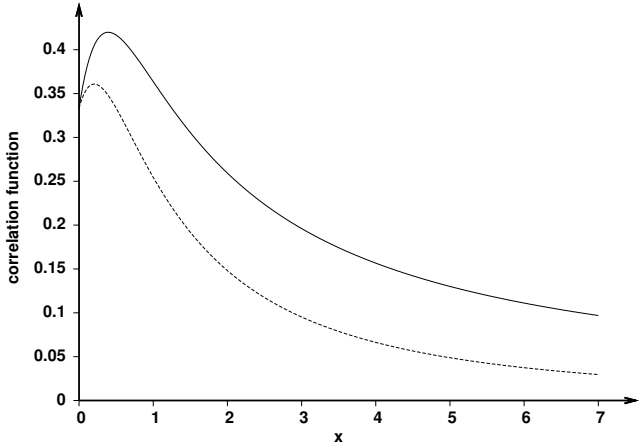}
\caption{The plot shows dependence of correlation functions
$\mathcal{C}^{\psi(k,k^\pi)}(\vec{a},\vec{b})$
(solid line) and  
$\mathcal{C}^{\psi(k,k^\pi)}_{\textsf{Cz}}(\vec{a},\vec{b})$ 
(dashed line) in the c.\ m.\ frame  
on $x$ for $\vec{a}\cdot\vec{b}=-1/2$, 
$\vec{a}\cdot\vec{n}=\vec{b}\cdot\vec{n}=1/2$.} 
\label{fig:bosons_correl}
\end{figure}

As one can check, the correlation function
$\mathcal{C}^{\Psi}_{\textsf{Cz}}(\vec{a},\vec{b})$ 
[Eq.~(\ref{eq:correl_func_CM_def})]
in the state (\ref{eq:state_bosons}) has the following form:
 \begin{multline}
 \label{eq:correl_func_bosons_Czachor_general}
 \mathcal{C}^{\psi(k,p)}_{\textsf{Cz}}(\vec{a},\vec{b}) \\
 = 2 \frac{-\vec{a}\cdot\vec{b}(kp)-
 (\vec{a}\cdot\vec{p})
 (\vec{b}\cdot\vec{k})}{\big(2+\frac{(kp)^2}{m^4} \big)
 \sqrt{m^2+(\vec{a}\cdot\vec{k})^2}
 \sqrt{m^2+(\vec{b}\cdot\vec{p})^2}}.
 \end{multline}
In the c.\ m.\ frame it reduces to
 \begin{multline}
 \label{eq:correl_func_bosons_Czachor_CMF}
 \mathcal{C}^{\psi(k,k^\pi)}_{\textsf{Cz}}(\vec{a},\vec{b}) \\
 = 2 \frac{-\vec{a}\cdot\vec{b}(1+2x) + x (\vec{a}\cdot\vec{n})
 (\vec{b}\cdot\vec{n})}{(2+(1+2x)^2)\sqrt{1+(\vec{a}\cdot\vec{n})^2x}
 \sqrt{1+(\vec{b}\cdot\vec{n})^2x}}.
 \end{multline}
In Fig.~\ref{fig:bosons_correl} we have plotted the functions
(\ref{eq:correl_func_bosons_our_CMF}) and
(\ref{eq:correl_func_bosons_Czachor_CMF}) for the fixed
configuration. We again observe local maxima for both functions.

\begin{figure}
\centering
\includegraphics[width=1\columnwidth]{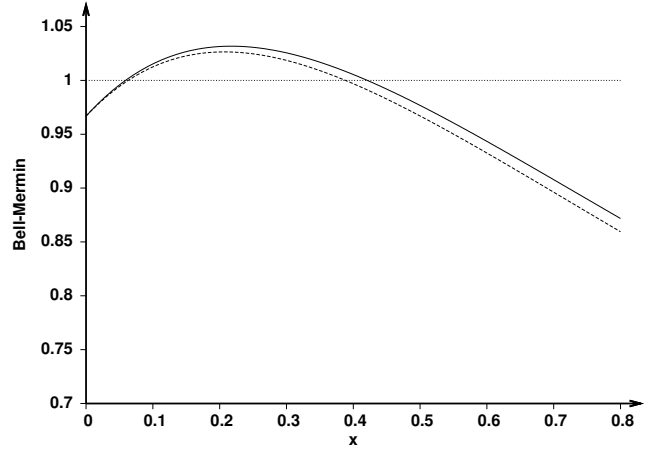}
\caption{The plot shows dependence of the left hand side of the
  Bell--Mermin inequality in the c.\ m.\
  frame for $\mathcal{C}^{\psi(k,k^\pi)}(\vec{a},\vec{b})$ 
  (solid line) and 
  $\mathcal{C}^{\psi(k,k^\pi)}_{\textsf{Cz}}(\vec{a},\vec{b})$
  (dashed line)
  on $x$ for $\vec{a}=(0.995004, 0, 0.0998334)$,
  $\vec{b}=(-0.40899, 0.907061, 0.0998334)$, $\vec{c}=(-0.581043,
  -0.807727, 0.0998334)$, and $\vec{n}=(0,0,1)$.}   
\label{fig:bosons_Bell_Mermin}
\end{figure}

In this case physical consequences are stronger than for $s=1/2$
particles.
According to the Mermin's paper \cite{cab_Mermin1980},
in the EPR-type experiments with the pair of spin 1 particles in the
singlet state
the following inequality has to be satisfied:
 \begin{equation}
  \textsf{Bell-Mermin}=C_{\vec{a}\vec{b}} + C_{\vec{b}\vec{c}} +
  C_{\vec{c}\vec{a}} \le1, 
  \label{eq:Bell_Mermin}
  \end{equation}
in the theory which fulfills the assumptions of local realism.
One can show \cite{cab_Mermin1980} that in the
nonrelativistic case this inequality is satisfied for each
configuration. However, both relativistic correlation functions
(\ref{eq:correl_func_bosons_our_CMF}) and
(\ref{eq:correl_func_bosons_Czachor_CMF}) can violate 
the inequality (\ref{eq:Bell_Mermin}). We have depicted such a
situation in Fig.~\ref{fig:bosons_Bell_Mermin}.

It should be stressed that previous works suggest that for fixed
measurements directions the degree of violation of Bell-type
inequalities by a pair of spin-1/2 particles monotonically decreases
with increasing velocity of the particles \cite{ALMH2003,TU2003_1}. 
Our results show that such a statement is false in general, also for
spin-1 particles. We have
shown that, at least for certain states, there exist configurations in
which the correlation functions and the degree of violation of
Bell-type inequalities have local extrema for some values of the
velocities of the EPR particles. Moreover, this strange behavior can
be observed 
for both discussed spin operators and for spin-1/2 as well as spin-1
particles. 
The most surprising fact is that EPR experiment in a fixed
configuration can distinguish the values of the velocity of the
particles corresponding to local extrema.
This observation is supported by
the recent results obtained in Ref.~\cite{Caban_2008_bosons_helicity},
where the helicity and linear polarization correlations of spin-1
particles  were analyzed. 

We have shown also that
relativistic quantum correlations are stronger then nonrelativistic
ones for a variety of configurations. Consequently, in such
configurations Bell inequalities are violated stronger by relativistic
correlations than by nonrelativistic ones.

Let us notice also that in some configurations the correlation
function and the degree of violation of CHSH inequality strongly
depend on the relativistic spin operator used in calculations
(compare e.g.\ Figs.~\ref{fig:fermions_CHSH_1} and
\ref{fig:fermions_CHSH_3}). 
This observation could help us to determine experimentally which of
the discussed spin operators is a proper one.
In the recent experiments with protons \cite{Sakai_etal2006} the
particles were too slow to distinguish different spin operators. 

The main result of our paper is the observation that the discussed
strange behavior of the correlation functions seems to be a general
property of the relativistic quantum mechanics
independent of the chosen relativistic spin operator.

\begin{acknowledgments}
This work has been supported by the University of Lodz grant and LFPPI
network.
\end{acknowledgments}


\end{document}